\documentstyle[preprint,psfig,aps]{revtex}
\draft
             
\begin{document}                % INITIALIZE - DONT CHANGE

\def\be{\begin{equation}}
\def\ee{\end{equation}}
\def\ba{\begin{eqnarray}}
\def\ea{\end{eqnarray}}
\def\ban{\begin{eqnarray*}}
\def\ean{\end{eqnarray*}}

%\preprint{DE-FG03-00-ER41132 - ???}

\title{Rotational spectra of weakly interacting Bose-Einstein condensates}
\author{Thomas Papenbrock\cite{tp} and George F. Bertsch\cite{gfb}}
\address{Institute for Nuclear Theory, Department of Physics, 
University of Washington, Seattle, WA 98195, USA}
\maketitle
\begin{abstract}
We study the spectrum of rotating Bose-Einstein Condensates in the limit of 
weak repulsive interactions and present analytical and numerical results for 
energies and wave functions. While the low-lying states are of collective 
nature, the high-lying states are dominated by single-particle excitations
and can accurately approximated by simple polynomial expressions.
In the limit that the number of particles is large compared to the number of 
excited quanta, the single-particle states become excellent approximations to
the eigenstates, and a rather simple ordering scheme is obtained.
\end{abstract}
The experimental realization of Bose-Einstein condensation in atomic vapors
\cite{Anderson,Bradley,Ketterle} has received much interest in recent years
\cite{Stringari}. Very recently experiments have started to investigate
Bose-Einstein condensates under rotation. Matthews {\it et al.} created
vortices in a two-component condensate \cite{JILA}, and Madison {\it et al.}
studied rotations of a one-component gas. Theoretical studies have focused on
the Thomas-Fermi regime of strong interactions \cite{Rokhsar,Linn,Feder} or on
the limit of weak interactions between the atoms
\cite{Wilkin,Mottelson,Cooper,Bertsch,Wilkin2,Kavoulakis,Jackson,Jackson2}.
While the former case is closer related to the experiment, the latter case
allows for exact (numerical or analytical) solutions and is of particular
theoretical interest. Wilkin {\it et al.} studied the case of attractive
interactions \cite{Wilkin} and employed a composite boson/fermion picture
\cite{Cooper,Wilkin2} to describe configurations beyond the one-vortex state in
few-body boson systems. Mottelson \cite{Mottelson} and the Copenhagen group
\cite{Kavoulakis,Jackson} analyzed the mean-field theory in the weak
interaction limit. In Ref.\cite{Bertsch} the authors studied the ground states
using exact diagonalization techniques. So far, research has been restricted to
the low-lying states in rotating Bose-Einstein condensates. It is the purpose
of this paper to study higher excited states in the rotational spectrum.

The Hamiltonian is 
\ban 
H=\sum_{j=1}^N \left\{-{1\over2}\nabla_j^2 + {1\over 2} {\bf r}_j^2\right\} 
+ g\sum_{i<j}2\pi\delta({\bf r}_i - {\bf r}_j),  
\ean 
where the trap frequency is set to $\hbar\omega=1$ and $g$ is a 
dimensionless coupling. In the limit of weak repulsive interactions $0<g\ll 1$ 
and maximal alignment of angular momentum with the $z$-axis the Hamiltonian
becomes essentially two-dimensional \cite{Mottelson,Kavoulakis}. In what
follows we restrict ourselves to this limit and fix the total angular momentum
to $L$. 

The problem then consists in diagonalizing the contact interaction in the
Hilbert space of degenerate oscillator states at energy total $L$ (setting 
the ground state energy to zero) \cite{Berry}. We will analyze the problem
in second quantization and in configuration space. While analytical results 
will be derived in both representations, numerical work is most easily done
using second quantization \cite{Haugset,PB}. Let us start with the second
quantized two-body interaction 
\be
\label{Hint}
\hat{V}=\sum_{i<j}2\pi\delta({\bf r}_i - {\bf r}_j)=
{1\over 2}\sum_{i,j,k,l}\frac{(k+l)!}{2^{k+l}\sqrt{i!j!k!l!}}\,\delta_{k+l}^{i+j}\,\,\hat{a}_i^\dagger \hat{a}_j^\dagger 
\hat{a}_k
\hat{a}_l.  
\ee 
The operators $\hat{a}_m$ and $\hat{a}_m^\dagger$ annihilate and create one
boson in the single-particle oscillator state $|m\rangle$ with energy
$m\hbar\omega$ and angular momentum $m\hbar$, respectively and fulfill bosonic
commutation rules. For total angular 
momentum $L$ the Fock space is spanned by states
$|\alpha\rangle\equiv|n_0,n_1,\ldots,n_k\rangle$ with $\sum_{i=0,k}n_i=N$,
$\hat{a}_j^\dagger\hat{a}_j|n_0,n_1,\ldots,n_k\rangle
=n_j|n_0,n_1,\ldots,n_k\rangle$ and $\sum_{j=0,k}j n_j=L$. Here $n_j$ denotes
the occupation of the $j^{\rm th}$ single particle state $|j\rangle$.

Turning to configuration space we 
use complex single-particle coordinates $z=x+iy$ such that $\langle
z|j\rangle = \phi_j(z)=(\pi j!)^{-1/2}\,z^j\,\exp{(-{1\over 2}|z|^2)}$.  Basis
functions for the many-body problem are
$\psi(z_1,\ldots,z_N)\prod_{j=1}^N\exp{(-{1\over 2}|z_j|^2)}$ where
$\psi(z_1,\ldots,z_N)$ is a homogeneous polynomial of degree $L$ that is 
totally
symmetric in its coordinates. For representations of such polynomials see,
e.g. ref.\cite{MacDonald}. In what follows we will omit the exponentials when
working in configuration space and identify wave functions
directly by their polynomial part. Let $\psi(z_1,\ldots,z_N)$ be such a
polynomial. It is easily seen that the two-body operator
\be
\label{op}
\hat{W}_{ij}=\sum_{m,n,k,l}\frac{(k+l)!}{2^{k+l}n!m!k!l!}\,\delta_{m+n}^{k+l}
\,z_i^n\,z_j^m\,\left\{\partial_{z_j}^k\partial_{z_i}^l\bigg|_{z_i=z_j=0}\right\}
\ee
has the same matrix elements as the contact interaction 
$2\pi\delta({\bf r}_i-{\bf r}_j)$. Here it is understood
that the derivatives act only onto the polynomial part of the wave function
and that they are evaluated at $z_i=z_j=0$. Performing the sums in
eq.~(\ref{op}) and introducing coordinates $R_{ij}=(z_i+z_j)/\sqrt{2}$ and 
$r_{ij}=(z_i-z_j)/\sqrt{2}$ leads to the following
\ba
\label{delta} 
2\pi\delta({\bf r}_i - {\bf
r}_j)\,\psi(z_1,\ldots,z_i,\ldots,z_j,\ldots,z_N) &=&
\hat{W}_{ij}\,\psi(z_1,\ldots,z_i,\ldots,z_j,\ldots,z_N)\nonumber\\
&=&\psi\left(z_1,\ldots,{z_i+z_j\over 2},\ldots,{z_j+z_i\over 2},
\ldots,z_N\right). 
\ea
Thus, the contact interaction $2\pi\delta({\bf r}_i-{\bf r}_j)$ acts on 
a wave function by replacing the $i^{\rm th}$ and $j^{\rm th}$ argument by
its mean value. Note that the center of mass
\ban
z_c={1\over N}\sum_{j=1}^N z_j
\ean
is left invariant under the action of the contact interaction.

Let us now consider the spectrum of states at angular momentum $L$.  States may
be classified by their angular momentum quantum numbers $L$ and $L_{\rm cm}$,
and by their energies $E_n=L+g\epsilon_n$. The angular momentum of the center
of mass and is denoted as $L_{\rm cm}$ and given by
\be
\label{lcm}
\hat{L}_{\rm cm}=z_c\sum_{j=1}^N\partial_{z_j}={1\over N}
\sum_{k}\sqrt{k+1}\,\hat{a}^\dagger_{k+1}\,\hat{a}_k\,\sum_{l}\sqrt{l+1}\,\hat{a}^\dagger_{l}\,\hat{a}_{l+1},
\ee 
where the differential operator is
understood to act only on the polynomial part of the wave function. The
functions $z_c$ and
\be
\label{q}
q_k(z_1,\ldots,z_N)=\sum_{j=1}^N(z_j-z_c)^k,\qquad k\ge 2
\ee
are eigenfunctions of $\hat{L}_{\rm cm}$ with eigenvalues one and zero,
respectively. Products $z_c^{\lambda_1}q_{\lambda_2}\ldots q_{\lambda_N}$ with
$\lambda$ being a partition of $L$ are eigenstates of $\hat{L}_{\rm cm}$ with
eigenvalue $L_{\rm cm}=\lambda_1$ and span the Hilbert space \cite{Berry} at
angular momentum $2\le L\le N$. Note that they do not form an orthogonal basis.
Let $|L,L_{\rm cm},\epsilon_n\rangle$ denote an eigenstate of the Hamiltonian
(\ref{Hint}). Due to an $SO(2,1)$ symmetry one has $z_c\,|L,L_{\rm
cm},\epsilon_n\rangle\propto|L+1,L_{\rm cm}+1,\epsilon_n\rangle$
\cite{Pitaevskii,Perelomov}. It is therefore of particular interest to find the
``non-spurious'' eigenstates (i.e. those with no excitations of the center of
mass coordinate) since all other states can simply be obtained by
multiplication with powers of $z_c$ \cite{Pitaevskii,Wilkin,Jackson2}.
Figure~\ref{fig1} (left part) shows the spectrum of non-spurious states for
$N=50$ bosons and angular momenta $2\le L\le 18$. Note that there are several
high-lying levels that approach constant values as $L$ increases. This is
particularly the case for two most energetic levels, and these are well
separated from their neighbors for all values of $L$. As has been observed
earlier in exact diagonalizations \cite{Bertsch} and mean-field calculations
\cite{Rokhsar,Kavoulakis}, the ground states depend linearly on total angular
momentum. 

Based on numerical results, we suggested in ref.\cite{Bertsch} that
the ground state wave function is  
\be
\label{psi0}
\psi_0(z_1,\ldots, z_N)=\sum_{1\le p_1<p_2<\ldots<p_L\le N} (z_{p_1}-z_c)
(z_{p_2}-z_c)\ldots(z_{p_L}-z_c),
\ee
and the ground state energy is
\be
\label{e0}
\epsilon_0 = {1\over 2}\left(N(N-1)-{NL\over 2}\right).
\ee
Very recently, Jackson and Kavoulakis \cite{Jackson2} showed the 
existence of an eigenstate with energy (\ref{e0}) while Smith and Wilkin 
\cite{Smith} proved that the wave function (\ref{psi0}) indeed has 
eigenvalue (\ref{e0}). This proof was given using the second quantized form
of the interaction. As an application of eq. (\ref{delta}) we now present a 
shorter proof in configuration space. We expand
\ban
\psi_0=\sum_{k=0}^L {N-k \choose L-k}\,(-z_c)^{L-k}\sum_{1\le
p_1<\ldots<p_k\le N} z_{p_1}\ldots z_{p_k}.
\ean
Applying operator (\ref{op}) to this expansion and using the prescription 
(\ref{delta}) yields  
\ban
\hat{W}_{ij}\,\psi_0=\sum_{k=0}^L {N-k \choose L-k}\,(-z_c)^{L-k}\Bigg\{&&
\sum_{1\le
p_1<\ldots<p_k\le N}^{\ne i,j}z_{p_1}\ldots z_{p_k}\nonumber\\
&&+(z_i+z_j)\sum_{1\le
p_1<\ldots<p_{k-1}\le N}^{\ne i,j}z_{p_1}\ldots z_{p_{k-1}}\\
&&+{1\over 4} (z_i+z_j)^2
\sum_{1\le
p_1<\ldots<p_{k-2}\le N}^{\ne i,j}z_{p_1}\ldots z_{p_{k-2}}\Bigg\}\nonumber.
\ean
Using $(z_i+z_j)^2=(z_i-z_j)^2+4z_iz_j$ in the third term enclosed in brackets 
one obtains 
\ban
\hat{W}_{ij}\,\psi_0=\psi_0+\sum_{k=0}^L {N-k \choose L-k}\,(-z_c)^{L-k}\left\{
{1\over 4} (z_i-z_j)^2
\sum_{1\le
p_1<\ldots<p_{k-2}\le N}^{\ne i,j}z_{p_1}\ldots z_{p_{k-2}}\right\}
\ean
Summing over $\sum_{i<j}$ and using the identity 
\ban
\lefteqn{\sum_{i<j}\left\{(z_i-z_j)^2\sum_{1\le
p_1<\ldots<p_{k-2}\le N}^{\ne i,j}z_{p_1}\ldots z_{p_{k-2}}\right\} 
=}&& \\  
&&-kN\sum_{1\le p_1<\ldots<p_k\le N} z_{p_1}\ldots z_{p_k}\,\,+\,\,
N(N-k+1)\,\,z_c\sum_{1\le p_1<\ldots<p_{k-1}\le N} z_{p_1}\ldots z_{p_{k-1}}
\ean
one arrives directly at the desired result $\hat{V}\psi_0=\epsilon_0\psi_0$.

Another exact result is readily obtained for $L=4$ and arbitrary $N$. 
Employing eq.~(\ref{delta}) one may show that 
\be
\label{L4}
\hat{V} \sum_{1\le i<j\le N}(z_i-z_j)^4 = {1\over 2}\left(N^2-{11\over 4} N +{3\over 2}\right)\,\,\sum_{1\le i<j\le N}(z_i-z_j)^4 
\ee
Unfortunately, we are not able to generalize this result
to higher values of angular momentum. Note however that wave functions
~(\ref{psi0}) and (\ref{L4}) comprise the set of non-spurious states for $L=4$.

Let us consider the most energetic non-spurious state. Numerical computations 
show that the ansatz (compare with eq.~(\ref{q}))
\be
\label{psi1}
\psi_1(z_1,\ldots,z_N)=q_L(z_1,\ldots,z_N)
\ee
has almost unit overlap with the most energetic non-spurious eigenstate and is
approaching the asymptotic eigenvalue $\epsilon_1={1\over 2}(N-1)(N-2)$ 
exponentially fast
with increasing values of angular momentum $L$. Table~\ref{tab1} presents
numerical values obtained for the overlaps with the exact numerical
eigenfunction, energies $\langle\psi_1|\hat{V}|\psi_1\rangle$, and widths 
$\Gamma_1^2=\langle\psi_1|\hat{V}^2|\psi_1\rangle-\langle\psi_1|\hat{V}|\psi_1\rangle^2$. The presented results show that eq. (\ref{psi1}) is a highly
accurate approximation to the exact eigenstate. This can be understood in more
detail by applying the operator (\ref{Hint}) to $\psi_1$
\be
\hat{V}\,\psi_1 = {1\over 2}(N-1)(N-2)\, \psi_1 + 
2^{1-L}\sum_{i<j}\left(z_i+z_j - 2z_c\right)^L.
\ee
For sufficiently large $L$ the ``remainder'' $\sum_{i<j}(z_i+z_j - 2z_c)^L$ is
numerically found to have significant overlap only with a small number of
eigenstates in the bulk. This explains the quality of our ansatz. Note that 
$\psi_0\propto\psi_1$ for $L=2$ and $L=3$. Note also that 
$\hat{V}\psi_1=[1+(-1)^L2^{1-L}]\psi_1$ for $N=3$. The ansatz
(\ref{psi1}) describes $\psi_1$ as an excitation of one particle relative to
the center of mass. This picture is confirmed by the computation of selected
single-particle occupation numbers $n_j=\langle\psi_1|\hat{n}_j|\psi_1\rangle$
shown in Figure~\ref{fig2}. The main contribution to the total angular momentum
stems from the single-particle orbital $\phi_L(z)$ and a smaller part is 
carried by the orbital $\phi_{L-1}(z)$. 
With view to eq.~(\ref{Hint}) it becomes clear that the
energy of $\psi_1$ depends very weakly on $L$. The energy contribution of the
highly excited single-particle states is exponentially suppressed for
sufficiently large angular momenta.

We may also obtain an excellent approximation for the second highest
non-spurious state. The ansatz
\be
\label{psi2}
\psi_2(z_1,\ldots,z_N)={N-1\over N}\,q_2(z_1,\ldots,z_N)\,
q_{L-2}(z_1,\ldots,z_N) - \psi_1(z_1,\ldots,z_N) 
\ee 
has almost unit overlap with the numerically obtained exact eigenstate for
sufficiently large values of total angular momentum. It approaches the 
numerically determined energy
$\epsilon_2={1\over 2}[(N-2)^2-1]$ exponentially fast with increasing $L$. 
Details concerning overlaps and width are presented in Table~\ref{tab1}.
The dominant contributions to the total angular
momentum of $\psi_2$ stem from the orbitals $\phi_2(z), \phi_{L-3}(z)$ and 
$\phi_{L-2}(z)$. As expected the
ansatz (\ref{psi2}) describes mainly a few-boson excitation.  The
single-particle character of the two most energetic states differs strongly
from the collective behavior found for the the ground state $\psi_0$.
\cite{Mottelson,Bertsch,Kavoulakis}. This observation is similar to the case 
of bosons in the Thomas-Fermi regime \cite{Guilleumas} and to the case of
many-fermion systems like atomic nuclei \cite{Bohr}.

We checked that the numerical results agree with ansatz (\ref{psi1}) and
(\ref{psi2}) for various numbers of bosons.  The convergence is
determined by $L$ only and depends weakly on $N$. The spectrum depicted in
Figure~\ref{fig1} seems to suggest that there are more states which may be
approximated by simple polynomial expressions; however we do not see an obvious
generalization of the expressions (\ref{psi1}) and (\ref{psi2}) to account for
further states.

Let us consider the regime $N\gg L\sim O(N^0)$ where the ground state is
occupied by a macroscopic number of bosons, i.e. $n_0\sim O(N)$. While 
low-lying states have been computed in refs.~\cite{Mottelson,Kavoulakis} we
aim at a general understanding of the full spectrum within this regime. 
Eq.~(\ref{Hint}) shows that the diagonal matrix elements of $\hat{V}$ are of
order $O(N^2)$ while off-diagonal ones are only of order $O(N^{1/2})$.  In a
first approximation one may then neglect any off-diagonal elements and take the
single-particle states as eigenstates. One obtains from eq.~(\ref{Hint}) \ba
\label{approx}
\hat{V}|n_0,n_1,n_2,\ldots,n_k\rangle &=& \epsilon_{\{n_j\}}|n_0,n_1,n_2,\ldots,n_k\rangle \,\,+ \,\,O(N^{1/2}), \nonumber\\
\epsilon_{\{n_j\}} &=& {1\over 2}\,n_0(n_0-1) 
+ 2n_0\sum_{k>0} 2^{-k}\,n_k\,.
\ea
Note that typical energies are of the order $O(N^2)$ while energy differences
for two different states are of order $O(N)$. Small energy differences may
occur for two configurations $\{n_j\}$ and $\{n'_j\}$ that differ only in the
occupation of the orbitals $j=0,2,3$. Both configurations should have identical
number of bosons and angular momentum, i.e. $n_0+n_2+n_3=n'_0+n'_2+n'_3$ and
$2n_2+3n_3=2n'_2+3n'_3$. The energy difference of such configurations is
only of order $\sim O(N^0)$ which is much smaller than a typical spacing. 
Thus there are quasi-degenerate levels in the spectrum, and the ground state
becomes quasi-degenerate for $L\ge 9$. This observation has also been made by
Mottelson~\cite{Mottelson}.  

Note that the single particle states have almost good quantum number $L_{\rm
cm}$. One finds $\hat{L}_{\rm cm}|n_0,n_1,n_2,\ldots,n_k\rangle\approx
n_1|n_0,n_1,n_2,\ldots,n_k\rangle$, and corrections are of order $O(L/N)$.
This can be seen by applying the operator~(\ref{lcm})
to a state with $n_0\sim O(N)$. Thus, the non-spurious states are comprised by
configurations with $n_1=0$. We may now discuss the order of the spectrum of
non-spurious states. According to eq.~({\ref{approx}) this ordering is led by
two observations. (i) The lower the number of bosons in the ground state, the
lower the total energy of this state. (ii) For two configurations with equal
numbers of excited bosons the higher energetic one has the highest single
particle orbital occupied.  This ordering scheme favors collective motion
above single particle excitations. Deviations from this scheme are caused by
the quasi-degeneracy of configurations discussed above. 
The validity of our arguments and of eq.~(\ref{approx}) is confirmed by
numerically computing and analyzing the spectrum for $N=46000$ in the regime
$2\le L\le 18$. 

Let us finally examine the $N$-dependence of the spectra at fixed $L$ as shown
in Figure~\ref{fig3}. The spectrum evolves rather smoothly with increasing $N$
and levels tend to be clustered at high $N$. For instance, there is a triplet 
of states at lowest energies. This is due to the
quasi-degeneracy discussed above, and the triplet is comprised by 
configurations
with $\{n_2=6\}$, $\{n_3=4\}$ and $\{n_2=3,n_3=2\}$. Such quasi-degeneracies 
indicate a high degree of integrability and regularity. However, there are also
a few quasi-degeneracies at mid-values of $N$. Closer inspection reveals that
these are avoided crossings. As an example we present in 
Figure~\ref{fig4}~(top)
the avoided crossing at $N\approx 70$. The enlargement of the energy
clearly indicates an avoided crossing. Further evidence comes from
analyzing the wave function structure, i.e. we follow the inverse 
participation \cite{Kaplan}
\be
I_\psi=\sum_\alpha|\langle\alpha|\psi\rangle|^4
\ee   
in the basis of single-particle basis states $|\alpha\rangle$. The inverse
$1/I_\psi$ measures the number of basis states that have significant overlap
with $|\psi\rangle$. Figure~\ref{fig4}~(bottom) shows that the wave function 
structure changes at the avoided crossing. The simple ordering obtained
for the spectrum at large values of $N$ is thus lost when decreasing $N$.
We also recall that avoided crossings indicate some degree of level repulsion 
and chaoticity \cite{GMW}. 
Therefore, we do not expect that all states may be described by simple 
polynomial expressions. Note however, that the high lying non-spurious states
and the ground state do not interact with their neighbored levels -- a further
confirmation of their regularity.

In summary, we have studied the rotational spectrum of Bose-Einstein
condensates in the limit of weak repulsive interactions. The most energetic
non-spurious states are well separated from neighboring levels and their
energies approach constant values with increasing angular momentum. These
states can accurately be approximated by simple polynomial expressions. They
exhibit single-particle behavior and differ thereby from the collective
behavior of the ground state.  In the limit of a macroscopic occupation of the
ground state and small total angular momentum the single-particle states become
approximate eigenstates. We obtain a simple ordering scheme for the entire
excitation spectrum that energetically favors collective behavior over
single-particle behavior. This ordering scheme is expected to break down when
leaving the limit of $L/N\ll 1$ since some states undergo avoided
crossings. While this introduces chaotic behavior in the involved states, the
ground state and the highest excited non-spurious states appear to remain
simple and regular. Parts of the analytical results presented in this work
including a proof concerning the ground state wave function are based on a
particularly useful and novel representation of the contact interaction in
configuration space.

We acknowledge conversations with J. Ginocchio and J. Verbaarschot and thank
K. Hagino for bringing ref.\cite{Berry} to our attention.  This work was
supported by the Dept. of Energy under Grant DE-FG03-00-ER41132.

\begin{table}
\begin{tabular}{|c||d|d|d|d|}
$L$                                          & 8 & 12 & 16 & 20 \\\hline\hline
$\langle\psi_1|\hat{V}|\psi_1\rangle$        & 1176.325  & 1176.019  &1176.001   &  1176.000 \\
$\Gamma_1$                                   & 0.6  &0.2   & 0.03   & 0.01  \\
$1-|\langle\psi_1|\psi_{\rm exact}\rangle|^2$& $2\times 10^{-4}$  & $6\times 10^{-6}$  & $2\times 10^{-7}$  & $< 10^{-7}$\\\hline
$\langle\psi_2|\hat{V}|\psi_2\rangle$        & 1153.031  & 1151.604  &1151.508   &  1151.501 \\
$\Gamma_2$                                   & 1.4  &0.3   & 0.09   & 0.03  \\
$1-|\langle\psi_2|\psi_{\rm exact}\rangle|^2$& $3\times 10^{-3}$  & $9\times 10^{-5}$  & $5\times 10^{-6}$  & $4\times 10^{-7}$\\
\end{tabular}
\protect\caption{Numerical data for the states $\psi_1$ and $\psi_2$ as a function of angular momentum $L$ for $N=50$. The energy expectation values $\langle\psi_j|\hat{V}|\psi_j\rangle$ approach energies $\epsilon_1=(N-1)(N-2)/2=1176$ and $\epsilon_2=((N-2)^2-1)/2=1151.5$, respectively. The widths $\Gamma_j^2=\langle\psi_j|\hat{V}^2|\psi_j\rangle-\langle\psi_j|\hat{V}|\psi_j\rangle^2$ are small compared to the level spacing (which is of order $O(N)$), and 
the overlap with the exact eigenstates is close to one.}  
\label{tab1}
\end{table}

\begin{figure}
  \begin{center}
    \leavevmode
    \parbox{0.9\textwidth}
           {\psfig{file=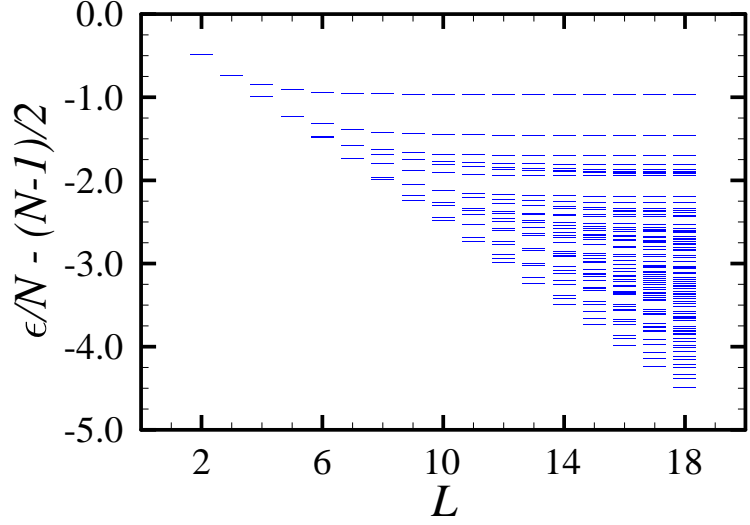,width=0.7\textwidth,angle=0}}
  \end{center}
\protect\caption{Spectrum of non-spurious states for a system of $N=50$ 
particles as a function of angular momentum.}
\label{fig1}
\end{figure}

\begin{figure}
  \begin{center}
    \leavevmode
    \parbox{0.9\textwidth}
           {\psfig{file=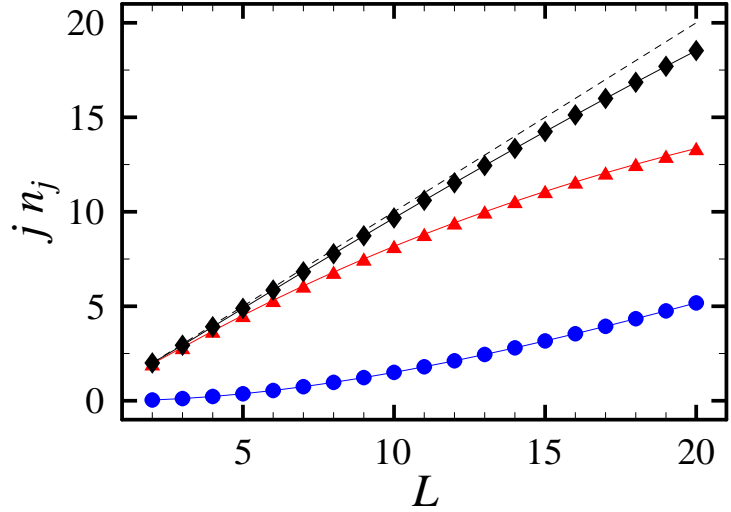,width=0.7\textwidth,angle=0}} 
  \end{center}
\protect\caption{Structure of state $\psi_1$: Single-particle angular 
momenta $j \langle\psi_1|\hat{n}_j|\psi_1\rangle$ as a function of 
angular momentum $L$ for a  system of $N=50$ bosons.($j=L-1$: circles; 
$j=L$: triangles; sum of both: diamonds; total angular momentum: dashed line)}
\label{fig2}
\end{figure}

\begin{figure}
  \begin{center}
    \leavevmode
    \parbox{0.9\textwidth}
           {\psfig{file=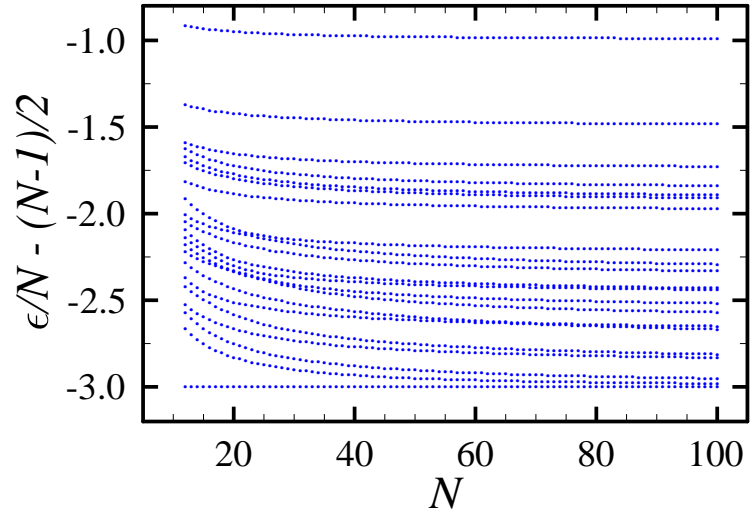,width=0.7\textwidth,angle=0}} 
  \end{center}
\protect\caption{Spectrum of non-spurious states for $L=12$ as a function
on the number of bosons $N$.}
\label{fig3}
\end{figure}

\begin{figure}
  \begin{center}
    \leavevmode
    \parbox{0.9\textwidth}
           {\psfig{file=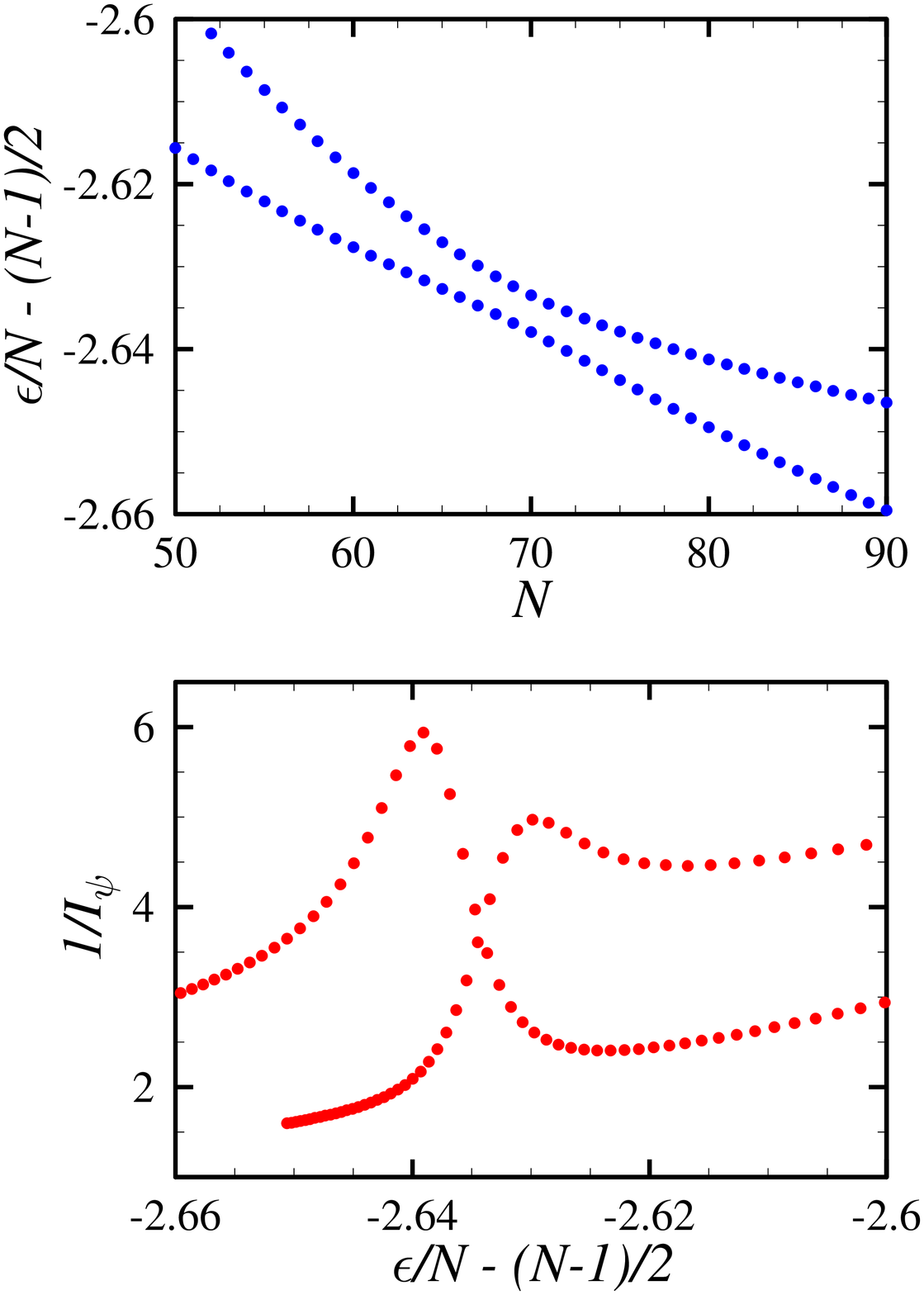,width=0.7\textwidth,angle=0}} 
  \end{center}
\protect\caption{Avoided Crossing. Top: Two energies as a function of $N$. Bottom: Number of participating basis states as a function of energy. Results are
taken from one avoided crossing at $L=12$.}
\label{fig4}
\end{figure}


\begin{references}  
%
\bibitem[a]{tp}
e-mail: papenbro@phys.washington.edu
%
\bibitem[b]{gfb}
e-mail: bertsch@phys.washington.edu
%
\bibitem{Anderson}
M. N. Anderson, J. R. Ensher, M. R. Matthews, C. E. Wieman, and E. A. Cornell,
Science {\bf 269}, 198 (1995).
%
\bibitem{Bradley}
C. C. Bradley, C. A. Sacket, J. J. Tollet, and R. G. Hulet,
\prl {\bf 75}, 1687 (1995).
%
\bibitem{Ketterle}
K. B. Davis, M.-O. Mewees, M. R. Andrews, N. J. van Druten, D. S. Durfee, 
D. M. Kurn, and W. Ketterle,
\prl {\bf 75}, 3969 (1995).
%
\bibitem{Stringari}
For a review see, e.g.,
F. Dalfovo, S. Giorgini, L. P. Pitaevskii, and S. Stringari,
\rmp {\bf 71}, 463 (1999).
%
\bibitem{JILA}
M. R. Matthews, B. P. Anderson, P. C. Haljan, D. S. Hall, C. E. Wieman, and
E. A. Cornell,
\prl {\bf 83} (1999) 2498.
%
\bibitem{Madison}
K. W. Madison, F. Chevy, W. Wohlleben, and J. Dalibard,
\prl {\bf 84}, 806 (2000).
%
\bibitem{Rokhsar}
D. A. Butts and D. S. Rokhsar,
Nature {\bf 397}, 327 (1999).
%
\bibitem{Linn}
M. Linn and A. L. Fetter,
\pra {\bf 60}, 4910 (1999).
%
\bibitem{Feder}
D. L. Feder, C. W. Clark, and B. I. Schneider,
\prl {\bf 82}, 4956 (1999);\pra {\bf 61}, 011601 (2000).
%
\bibitem{Wilkin}
N. K. Wilkin, J. M. Gunn, and R. A. Smith,
\prl {\bf 80}, 2265 (1998).
%
\bibitem{Mottelson}
B. Mottelson,
\prl{\bf 83}, 2695 (1999).
%
\bibitem{Cooper}
N. R. Cooper and N. K. Wilkin,
\prb {\bf 60}, R16279 (1999).
%
\bibitem{Bertsch}
G. F. Bertsch and T. Papenbrock,
\prl {\bf 83}, 5412 (1999).
%
\bibitem{Wilkin2}
N. K. Wilkin and J. M. F. Gunn,
\prl {\bf 84}, 6 (2000).
%
\bibitem{Kavoulakis}
G. M. Kavoulakis, B. Mottelson, and C. J. Pethick,
arXiv:cond-mat/0004307.
%
\bibitem{Jackson}
A. D. Jackson, G. M. Kavoulakis, B. Mottelson, and S. M. Reimann,
arXiv:cond-mat/0004309.
%
\bibitem{Jackson2}
A. D. Jackson and G. M. Kavoulakis,
arXiv:cond-mat/0005159.
%
\bibitem{Haugset}
T. Haugset and H. Haugerud,
\pra {\bf 57}, 3809 (1998).
%
\bibitem{PB}
T. Papenbrock and G. F. Bertsch,
\pra {\bf 58}, 4854 (1998).
%
\bibitem{Berry}
G. Natanson and R. S. Berry,
Ann. Phys. {\bf 155}, 158 (1984); Ann. Phys. {\bf 155}, 178 (1984).
%
\bibitem{MacDonald}
I. G. Macdonald,
{\it Symmetric Functions and Hall Polynomials}, 2nd edition, Clarendon Press, 
Oxford 1995.
%
\bibitem{Pitaevskii} 
L.P. Pitaevskii and A. Rosch, 
\pra {\bf 55}, R853 (1998).
%
\bibitem{Perelomov}
A. Perelomov, 
{\it Generalized Coherent States and Their Applications},
(Texts and monographs in physics), Springer Verlag, Berlin 1986. 
%
\bibitem{Smith}
R. A. Smith and N. K. Wilkin,
arXiv:cond-mat/0005230.
%
\bibitem{Guilleumas}
F. Dalfovo, S. Giorgini, M. Guilleumas, L. Pitaevskii, S. Stringari,
\pra {\bf 56}, 3840 (1997). 
%
\bibitem{Bohr}
N. Bohr and B. R. Mottelson,
{\it Nuclear Structure}, Vol. I (Benjamin, New York, 1969).
%
\bibitem{Kaplan}
See, e.g.
L. Kaplan, 
Nonlinearity {\bf 12}, R1 (1999).  
%
\bibitem{GMW}
T. Guhr, A. M\"uller-Groeling, and H. A. Weidenm\"uller,
Phys. Rep. {\bf 299}, 189 (1998).
%
\end{references}
\end{document}